\documentclass[11pt,a4paper]{article}
\usepackage[margin=1in]{geometry}
\usepackage[utf8]{inputenc}
\usepackage{amsmath,amssymb,amsfonts}
\usepackage{graphicx}
\usepackage{authblk}
\usepackage{xcolor}
\usepackage{siunitx}
\usepackage{booktabs}
\usepackage{tabularx}
\usepackage{hyperref}
\usepackage{threeparttable}
\usepackage{biblatex}
\newcommand{\NaIrCl}{Na$_2$IrCl$_6$}

\title{Na$_2$Ir$^\text{IV}$Cl$_6$: The Missing Member of the Perfectly Cubic Vacancy Ordered $A_2$IrCl$_6$ Family, Another Potential $J_{eff}$~=~$\frac{1}{2}$ Ground State Candidate}

\author[1]{Siddhartha Sankar Soren}
\author[1]{Aditya Patra}
\author[2]{Rounak Das}
\author[2]{Ritwik Das}
\author[3]{Tanmay Chanda}
\author[4]{Jens Buck}
\author[2]{Indra Dasgupta}
\author[1]{Sugata Ray \thanks{Corresponding author: mssss2407@iacs.res.in}}

\affil[1]{\textit{School of Materials Sciences, Indian Association
for the Cultivation of Science, Kolkata 700032, India}}

\affil[2]{\textit{School of Physical Sciences, Indian Association
for the Cultivation of Science, Kolkata 700032, India}}

\affil[3]{\textit{School of Chemical Sciences, Indian Association
for the Cultivation of Science, Kolkata 700032, India}}

\affil[4]{\textit{Ruprecht Haensel Laboratory, Deutsches Elektronen--Synchrotron
	DESY, 22603 Hamburg, Germany; Institut f\"ur Experimentelle und
	Angewandte Physik, Christian-Albrechts-Universit\"at zu Kiel,
	24098 Kiel, Germany}}

\date{\today}

\begin{document}

\maketitle

\begin{abstract}
Vacancy ordered double perovskites (VODPs) have been known for more than 100 years, but their exact crystal structures and magnetic ground state remained controversial till date. Anhydrous VODP iridates ($A_2$Ir$X_6$; $A$~=~K$^+$, (NH$_4$)$^+$, Cs$^+$, Rb$^+$, and $X$~=~F$^-$, Cl$^-$, Br$^-$), having {\it face-centered cubic} (fcc) antiferromagnetic ground state, offer a chemically well-defined platform to investigate the influence of on-site Coulomb repulsion, crystal field, spin-orbit coupling, and competing Heisenberg and Kitaev interactions. Interestingly, among these, the Na-based VODP iridates seem to possess an unusual hydration dependent structural and magnetic uncertainty. As a result, a cubic Na$_2$IrCl$_6$ system having fcc antiferromagnetism could not be stabilized ever. Here, for the first time, we report the synthesis and comprehensive characterization of anhydrous Na$_2$IrCl$_6$ single crystals having fcc structure with antiferromagnetic transition temperature of 1.91~K. Thermogravimetric analysis establishes the absence of lattice water and also demonstrates thermal stability of the crystals up to 453 K. The cubic phase reported here contrasts with previously described 'anhydrous' monoclinic Na$_2$IrCl$_6$ and highlights the strong structural and physical property relationship in Na-based VODP iridates. This new material offers an exciting playing ground for exploring another potential perfect $J_{eff}$~=~$\frac{1}{2}$ octahedra and the consequent strong Kitaev interaction within the nearest-neighbor (nn) pairs of Ir ions. 
\end{abstract}

\section{INTRODUCTION}

Perfectly cubic antifluorite-like vacancy ordered double perovskite (VODP) from $A_2$IrCl$_6$ family ($A$~=~K, NH$_4$, Rb, and Cs monovalent cations) have been studied for decades as a model {\it face-centered cubic} (fcc) antiferromagnet, possessing long-range superexchange pathways between Ir$^{4+}$ ions mediated by the strongly covalently-linked halogen ions (see Fig. 1 and the shaded area highlighting the Ir$^{4+}$-Cl-Cl-Ir$^{4+}$ connectivity)~\cite{r32}. These systems have resurfaced in recent times when honeycomb magnets with $J_{eff}$~=~$\frac{1}{2}$ ground state have been vehemently chased in search of most plausible realization of bond-anisotropic Kitaev physics which may give rise to quantum spin liquid states, characterized by unusual spin dynamics with fractionalized Majorana fermions~\cite{r48}, even though the fcc structure taken up by the title compounds does not seem compatible with the coveted honeycomb symmetry. However, if the Ir$^{4+}$ ions having 5$d^5$ electronic configuration indeed settles down within a perfect octahedral environment, as has been argued in various works on this series of compounds~\cite{r49,r50,r51,r52,r53,r54,r55,r56,r34,r25,r57,r26,r59,r60}, the coveted $J_{eff}$~=~$\frac{1}{2}$ ground state could be realized, and interestingly it has also been predicted~\cite{r54,r61,r62,r63} that these $J_{eff}$~=~$\frac{1}{2}$ states even within a fcc framework may host the bond-directional (Kitaev) anisotropy through the 90$^{\circ}$ Ir–O–Ir superexchange. It might also be worth noting here that not only in iridates, even in K$_2$RuCl$_6$, the 4$d$ analog of the 5$d^4$ system (Ir$^{5+}$), the unquenched orbital moment of Ru ($L$~=~1) within the perfect [RuCl$_6$]$^{2-}$ octahedra helps to realize an unusual $J_\text{eff}$ = 0 ground state~\cite{r36}, as recently probed using Ru $L_3$-edge resonant inelastic x-ray scattering (RIXS) experiments. However, coming back to the discussion of VODP iridates, firstly there had been controversies regarding the realization of true $J_{eff}$~=~$\frac{1}{2}$ ground states in these compounds~\cite{r25,r26,r59}, especially an observed splitting in a loss feature at 0.6 eV of the RIXS spectra have been interpreted in contradicting ways with a strong prediction of local octahedral distortions and lifting of $J_{eff}$~=~$\frac{1}{2}$ state degeneracy. On the other hand, despite the prediction of Khan {\it et al.}~\cite{r34} that other than a weak next-nearest-neighbor (nnn) exchange ($J_2~>$~0~\cite{r64}), coexistence of strong Heisenberg exchange $J_1~>$~0 and Kitaev interaction $K~>$~0 for the nearest-neighbor (nn) pairs of Ir$^{4+}$ ions is also essential to explain the fcc antiferromagnetism, Bhaskaran et al.~\cite{r53} concluded that the Kitaev term in (NH$_4$)$_2$IrCl$_6$ and K$_2$IrCl$_6$ is negligibly small compared to Heisenberg interaction strength ($J_1~\gg~K$) in these series of compounds. However, a recent work interestingly predicted realization of nodal line spin liquid in K$_2$IrCl$_6$~\cite{r60} as an outcome of ideal realization of the Heisenberg-Kitaev model on the fcc lattice with enhanced quantum fluctuations. In short, the controversy and quest for ideal [Ir$X_6$]$^{2-}$ octahedra within perfectly fcc $A_2$IrCl$_6$ antiferromagntetic symmetry continue to persist even after six decades of active discussion~\cite{r24}.

    \begin{figure}[!ht]
\centering
\includegraphics[width=0.87\textwidth]{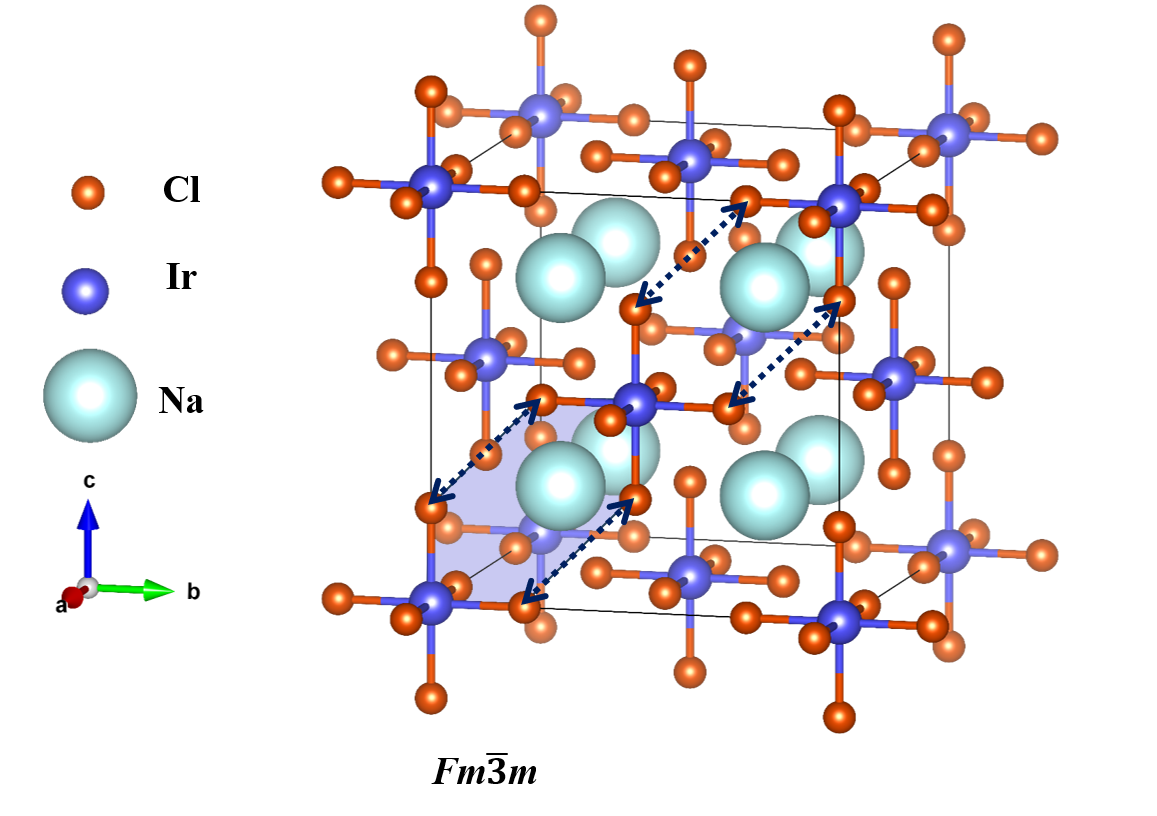}
\caption{Schematic of Na$_2$IrCl$_6$ cubic crystal structure and super exchange pathway between nearest neighbour magnetic Ir$^{4+}$ ions.}
\label{fgr:Introduction}
\end{figure}

Surprisingly, in all these decades of study, the absence of cubic Na$_2$IrCl$_6$ in the scenario has not attracted much attention. It appeared that particularly the Na-member of the series remains severely vulnerable against water inclusion within the crystal structure, making the stabilization of the expected cubic symmetry untenable. Bao \textit{et al.} comprehensively investigated this system, which was readily hydrated to form two stable phases; Na$_2$IrCl$_6$.2H$_2$O and Na$_2$IrCl$_6$.6H$_2$O~\cite{r29}. Interestingly, the author also claimed to have synthesized anhydrous Na$_2$IrCl$_6$ which unexpectedly stabilizes within a monoclinic $P$2$_1$/$n$ structure and is a canted antiferromagnet with an ordering temperature of 7.4 K. The reason why the anhydrous Na$_2$IrCl$_6$ got stabilized in a monoclinic structure, defying all expectations, however remained unexplained. It should be noted that a recent theoretical work~\cite{r30} on hypothetical anhydrous cubic Na$_2$IrCl$_6$ predicted that unlike other similar members of the series, this system, if exists at all, would prefer ferromagnetic halfmetalicity over fcc antiferromagnetism.\\
In this work, for the first time, we report the synthesis of the anhydrous single crystal of Na$_2$IrCl$_6$, which consistently crystallizes in cubic structure with $Fm\Bar{3}m$ space group. Our single crystal X-ray diffraction refinement confirms that Na$_2$IrCl$_6$ remains cubic at least down to 165 K. Thermogravimetric measurements confirm the anhydrous nature of our single crystals. X-ray photoelectron spectroscopy and $dc$ magnetization measurements show long range antiferromagnetic order of the Ir$^{4+}$ ions at 1.91~K in this system, which is consistent with other cubic members of the $A_2$IrCl$_6$ family. Certain modifications of the synthesis conditions made it possible to grow this crystal for the first time, which should offer a new  exploration stage for studying the `perfect' $J_{\mathrm{eff}}$~=~$\frac{1}{2}$ ground state and also to probe the applicability of the ideal Heisenberg-Kitaev model in this class of systems.

\section{EXPERIMENTAL SECTION}
\subsection{Material Synthesis}
Commercially available Na$_2$IrCl$_6$.6H$_2$O powder (99.99\% purity, Alfa Aesar) was readily dissolved in Milli-Q water in a glass beaker. Then beaker was kept inside a fume hood in a air-conditioned lab maintaining the room temperature at 24 \textdegree C in Indian humidity condition for a week. Due to slow evaporation of water, several black octahedral or hexagonal-shaped crystals of Na$_2$IrCl$_6$ with a diameter of 1-2 mm were obtained at the bottom of the glass beaker. The total duration of crystal formation either increase in summer due to high humidity or decrease in winter due to low humidity. For structural and magnetic comparison, K$_2$IrCl$_6$ single crystals were also grown using similar procedure. The present method of synthesis of single crystals is quiet different from the earlier reported procedures of K$_2$IrCl$_6$ or (NH$_4$)$_2$IrCl$_6$ where the anhydrous salts were dissolved in distilled water with slightly acidic condition. Then solutions are kept in furnace box with temperature varing from 60 \textdegree C to 90 \textdegree C for 10hrs to several days and cooled down to room temperature with controlled slow evaporation over 2days to a week \cite{r25,r57,r28,r66}. There is no earlier report available on any attempt to synthesize Na$_2$IrCl$_6$ single crystal from any solvent evaporation method. Even Bao \textit{et al.} synthesized Na$_2$IrCl$_6$ polycrystal from Na$_2$IrCl$_6$.6H$_2$O powder by heating at 200 \textdegree C in vaccumm \cite{r29}.

\subsection{X-ray Single Crystal and Powder Diffraction Measurement}

\begin{figure}[!ht]
\centering
\includegraphics[width=0.87\textwidth]{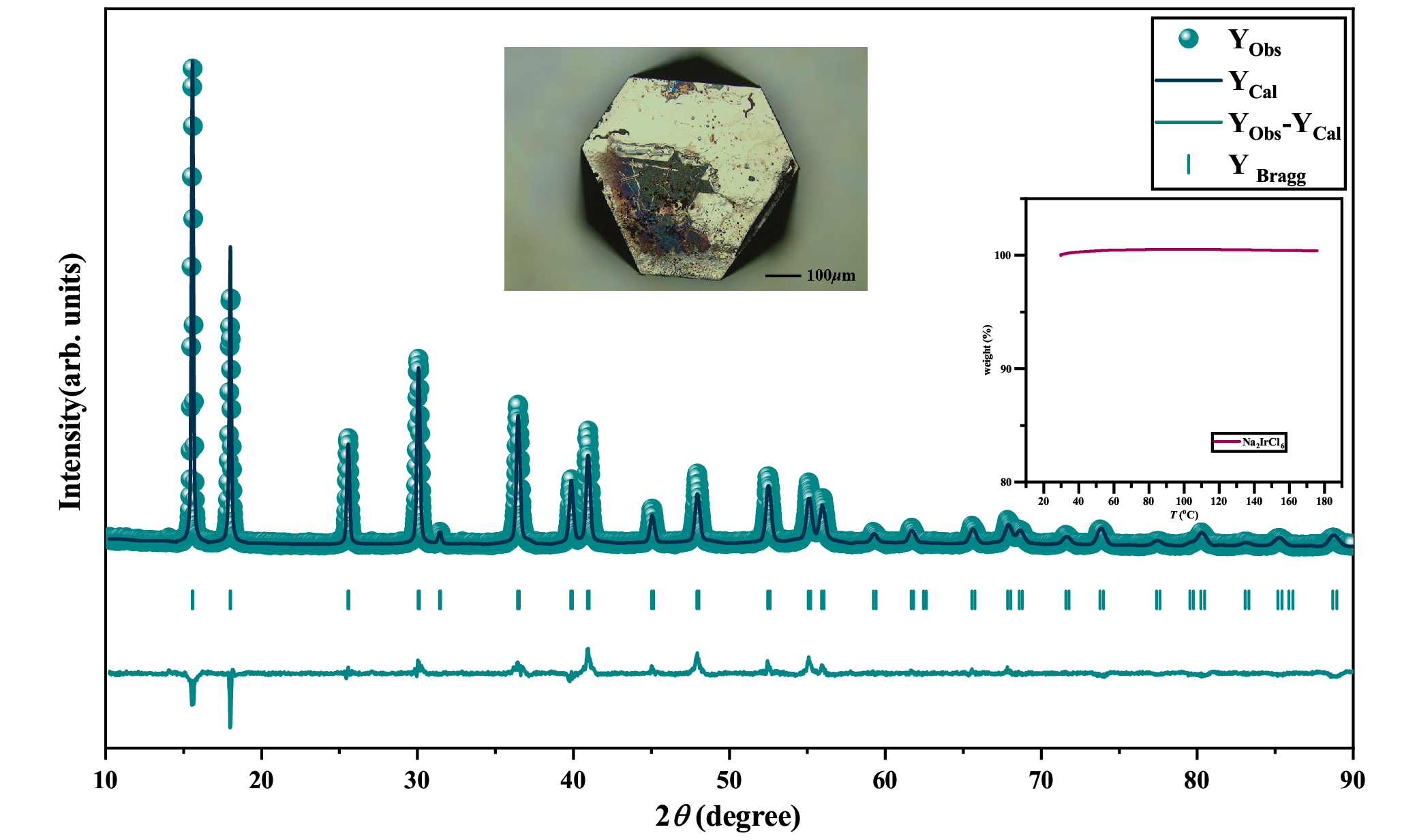}
\caption{Experimental and refined Xray powder diffraction patterns with Bragg reflections. Inset: Thermogravimetric analysis of Na$_2$IrCl$_6$ heating from 30\textdegree-180\textdegree C at a rate of 10\textdegree C/min in dry nitrogen environment.}
\label{fgr:XRD}
\end{figure}

\begin{table}[!ht]
\caption{Crystallographic and Experimental Data and Structure Refinement Parameters for a Na$_2$IrCl$_6$ single crystal}
\centering
\begin{tabular}{cc}
\toprule
\textbf{Empirical formula} & \textbf{Na$_2$IrCl$_6$} \\
\midrule
Molecular weight & 450.91 \\
Temperature (K) & 165 \\
Crystal system & Cubic \\
Wavelength (Å) & 0.71073 \\
Space group & $Fm\Bar{3}m$ \\
$a$ (\AA) & 9.799(3) \\
$V$ (\AA$^3$) & 941.0(5) \\
$Z$ & 4 \\
Number of measured reflections & 1314 \\
Number of independent reflections & 63 \\
Number of reflections with $I > 2$$\sigma$($I$) & 63 ($R_{int}= 0.0714$) \\
$\theta$ (deg) & 50.038 \\
Absorption coefficient ($\mu$) (mm$^{-1}$) & 15.90 \\
Number of specified parameters & 7 \\
$R$ ($I > 2$$\sigma$($I$)) & 0.0285 \\
$wR$ ($I > 2$$\sigma$($I$)) & 0.0710 \\
GOOF & 1.292 \\
\bottomrule
\end{tabular}
\label{tab:SCXRD_FITTING}
\end{table}

The phase identification and purity of the powder prepared from single crystal was verified from the X-ray powder diffraction (XRPD) patterns obtained on a Bruker D8 Advance diffractometer (Vantec position-sensitive detector, Ge monochromatized Cu $K_\alpha$ radiation, Bragg-Brentano geometry, DIFFRACT plus software) in the 2$\theta$ range of 10\textdegree-90\textdegree~~with a step size of 0.02\textdegree~~(counting time was 15 sec per step). XRPD pattern (Figure~\ref{fgr:XRD}) was analyzed with the Rietveld method implemented in the FULLPROF software. The diffraction peaks were described by a thompson-cox hastings pseudo-voigt profile function. A peak asymmetry correction was made for angles below 40\textdegree~~(2$\theta$). Background intensities were estimated by linear interpolation between 60-70 selected points.\\
The experimental X-ray structural data set for the Na$_2$IrCl$_6$ single crystal was obtained on a Bruker SMART APEX II automatic diffractometer equipped with an Oxford Cobra low-temperature cryosystem at 165 K (graphite monochromatic Mo $K_\alpha$ radiation, $\lambda$ = 0.71073 \AA, $\omega$-scan). Integration of X-ray diffration peak intensities was carried out using the $SAINT$ version 8.38A software; the $SADABS$-2016 program included in the $APEX$-3 $SMART$ software package~\cite{r44} was used to take into account absorption by crystal habitus, the reduction to a uniform scale, and the averaging of the integral intensities of the experimental data. Structural refinements were done using $SHELX$-2014 software~\cite{r45,r46} using the XPREP utility to determine the correct space group. The model of crystal structure was refined by full-matrix least-squares over $F$$^2$$_{hkl}$ with anisotropic thermal parameters for all atoms. The final refinement, treatment of disorder, and finalization of CIF were carried out in $OLEX2$, version 1.2.8. The parameters of the X-ray diffraction experiment and the final results of crystal structure refinements are presented in Table~\ref{tab:SCXRD_FITTING}.\\

\subsection{Thermogravimetric Measurement}
Thermogravimetric analysis was performed using a TA perkinelmer TGA 4000. Na$_2$IrCl$_6$ single crystal was heated to 180\textdegree C at a rate of 10\textdegree C/min in dry nitrogen environment (20 mL/min). It was then kept at the temperature for 2 min before being cooled to 30\textdegree C. The temperature was not taken any higher as the sample dissociates at higher temperatures. 

\subsection{X-ray photoelectron spectroscopy (XPS) Measurement}

X-ray photoelectron spectroscopy (XPS) experiments have been carried out at the P04 beamline of the Deutsches Elektronen-Synchrotron DESY in Germany (Proposal No. I-20251160) using an (energy = 1000 eV) monochromatic x-ray source with an energy resolution of 0.1 eV. The surface of the single crystal was cleaved using a pin $in$ $situ$ chamber. The collected spectra was then processed and analyzed with the KOLXPD software, after providing a shift of 10.64 eV, coming due to charging effect.

\subsection{Magnetic Measurement}

Dc magnetization was measured in a superconducting quantum interference device (SQUID, Quantum Design) magnetometer, MPMS3. The $dc$ magnetic susceptibility $\chi(T)$ of Na$_2$IrCl$_6$ was measured under an applied field of 500 Oe in a temperature range from 1.8 to 300 K. The isothermal magnetization $M$($H$) measurement was collected by SQUID with a field of up to 7 T.

\subsection{Theoretical Calculation}

All the electronic structure calculations are carried out using DFT in the pseudopotential plane-wave basis within the generalized gradient approximation \cite{PBE1996} supplemented with Hubbard $U$ \cite{HubbardU1991} as encoded in the Vienna $ab$ $initio$ simulation package (VASP) \cite{abinitio1993, abinitio1996} with projector augmented wave potentials \cite{PAW1994, pseudopotential1999}. The calculations are done with standard values of $U_d = 1.7$ eV and $J_H = 0.3$ eV \cite{Valenti_2016} chosen for Ir. We use the energy cutoff for the plane-wave basis as 600 eV and a $8 \times 8 \times 8$ k-point grid for the primitive unit cell of Na$_2$IrCl$_6$. Maximally localized Wannier functions (MLWFs) were constructed using WANNIER90 to extract the nonmagnetic hopping parameters \cite{WAN_1, WAN_2}. Exchange parameters are calculated using projED method \cite{Valenti_2016, Dasgupta_2026, Das2025} as well as using perturbation theory \cite{Valenti_2016}.   

\section{RESULT AND DISCUSSION}

\subsection{Crystal Structures}

\begin{table}[!ht]
\caption{Room temperature crystallographic information of Na$_2$IrCl$_6$: cubic phase of space group $Fm\Bar{3}m$ (space group no. 225, $a$ = $b$ = $c$ = 9.857(6) \AA, $\alpha$ = $\beta$ = $\gamma$ = 90\textdegree) is used to refine the powder XRD pattern. Na$_2$IrCl$_6$: $R_\text{p}$ = 19.5, $R_\text{wp}$ = 19.0, $R_\text{exp}$ = 9.96, $\chi^2$ = 3.64.}
\centering
\begin{tabular}{cccccc}
\toprule
\textbf{Atom} & \textbf{\textit{x}} & \textbf{\textit{y}} & \textbf{\textit{z}} & \textbf{Occupancy} & \textbf{\textit{B}(\AA)$^2$} \\
\midrule
Na & 0.25 & 0.25 & 0.25 & 1.00 & 8.13(2) \\
Ir & 0.00 & 0.00 & 0.00 & 1.00 & 0.12(3) \\
Cl & 0.2456(0) & 0.00 & 0.00 & 1.00 & 0.94(8) \\
\bottomrule
\end{tabular}
\label{tab:XRD_FITTING}
\end{table}

Powder XRD as well as single crystal diffraction data of Na$_2$IrCl$_6$ crystals both reveal that the grown crystals belong to the hitherto unknown class of fcc crystals. The system is clearly isostructural (fcc, space group $Fm\Bar{3}m$, $Z$ = 4, $a$ = 9.762(2) \AA) with the known K$_2$IrCl$_6$ crystal~\cite{r47} and consists of discrete [IrCl$_6$]$^{2-}$ complex anions and Na$^+$ cations. The anion occupies a special position with the $O_h$ symmetry ($m$$\Bar{3}$$m$) and hence is a regular octahedron with Ir–Cl distances of 2.420(3) \AA. The Na$^+$ cation occupies a special position with $T_d$ symmetry ($\Bar{4}$3$m$) and is surrounded by 12 Cl atoms with the Na-Cl distances of 3.485(4) \AA. Table~\ref{tab:XRD_FITTING} presents the lattice parameters, atomic positions, site occupancy, and goodness factors determined by the Reitveld refinements. It is important to note here that despite a smaller size of the Na$^+$ ion compared to K$^+$, the Na$_2$IrCl$_6$ crystallizes with a larger unit cell volume~\cite{r25,r31,r39} (see Table~\ref{tab:ir_table}). More interestingly, despite the fact that the Ir-Cl bond lengths within the [IrCl$_6$]$^{2-}$ octahedra in Na$_2$IrCl$_6$ is lower by 1.3\% than that in K$_2$IrCl$_6$~(see Table~\ref{tab:ir_table}), the Cl-Cl distance appears longer by 3.3\%, making the overall unit cell larger in the sodium system. This probably indicates that the Ir-Cl covalency~\cite{r32} is relatively larger in K$_2$IrCl$_6$ than its sodium counterpart, making the negative charge density on Cl$^-$ and consequently, inter-ion repulsion higher in Na$_2$IrCl$_6$.\\

\begin{table}[!ht]
\centering
\caption{Comparison of Crystal Structures and Magnetic Properties in Na$_2$IrCl$_6$ and Related Ir$^{\text{IV}}$ Hexahalides.}
\label{tab:ir_table}
\renewcommand{\arraystretch}{1.15}
\begin{threeparttable}
\begin{tabularx}{\textwidth}{@{}l *{3}{>{\centering\arraybackslash}X} >{\raggedright\arraybackslash}p{3cm} >{\raggedright\arraybackslash}p{2cm}@{}}
\toprule
 & \textbf{Ionic radii of A ion in 12 co-ordination (\AA)} & \textbf{Lattice parameter (\AA)} & \textbf{Ir-Cl bond distance (\AA)} & \textbf{Cl-Cl superexchange distance (\AA)} & \textbf{$T_\text{N}$ (K)} \\
\midrule
K$_2$IrCl$_6$ & 1.64 & 9.762~\cite{r47} & 2.451 & 3.436 & 3.08 \\
(NH$_4$)$_2$IrCl$_6$ & 1.67 & 9.870~\cite{r69} & 2.470 & 3.49 & 2.15 \\
Na$_2$IrCl$_6$ & 1.39 & 9.858~\cite{r47} & 2.420 & 3.551 & 1.91 \\
Rb$_2$IrCl$_6$ & 1.83 & 9.916~\cite{r68} & NA & NA & 1.85 \\
Cs$_2$IrCl$_6$ & 1.88 & 10.212~\cite{r67} & 2.330 & 3.92 & Less than 1.6 \\
\bottomrule
\end{tabularx}
\end{threeparttable}
\end{table}

\subsection{Thermogravimetric Analysis}

We believe that the anhydrous monoclinic Na$_2$IrCl$_6$ reported in ref~\cite{r29} was not completely anhydrous and therefore could not be stabilized within the cubic crystal structure, while we have been able to successfully grow the completely anhydrous version of the material. In order to confirm this, thermogravimetric measurements have been carried out with the crystals (inset of Figure~\ref{fgr:XRD}) which revealed that almost no water is present in the present set of Na$_2$IrCl$_6$ single crystals. Clearly, neither any weight loss could be seen by heating the crystals upto 180\textdegree~C (453 K) nor the system gained any mass by consequently cooling it down to room temperature (not shown). From this, it can be concluded that Na$_2$IrCl$_6$ single crystals shown here  are completely anhydrous in nature. In order to confirm that the sample structure remained stable after heating upto 180\textdegree~C, single crystal x-ray diffraction was remeasured and it was confirmed that the sample remained stable in the cubic $Fm\Bar{3}m$ space group. The parameters of single crystal structure refinements before and after TGA are given in the Supporting Information~\cite{r47}

\begin{figure}[!ht]
\centering
\includegraphics[width=0.87\columnwidth]{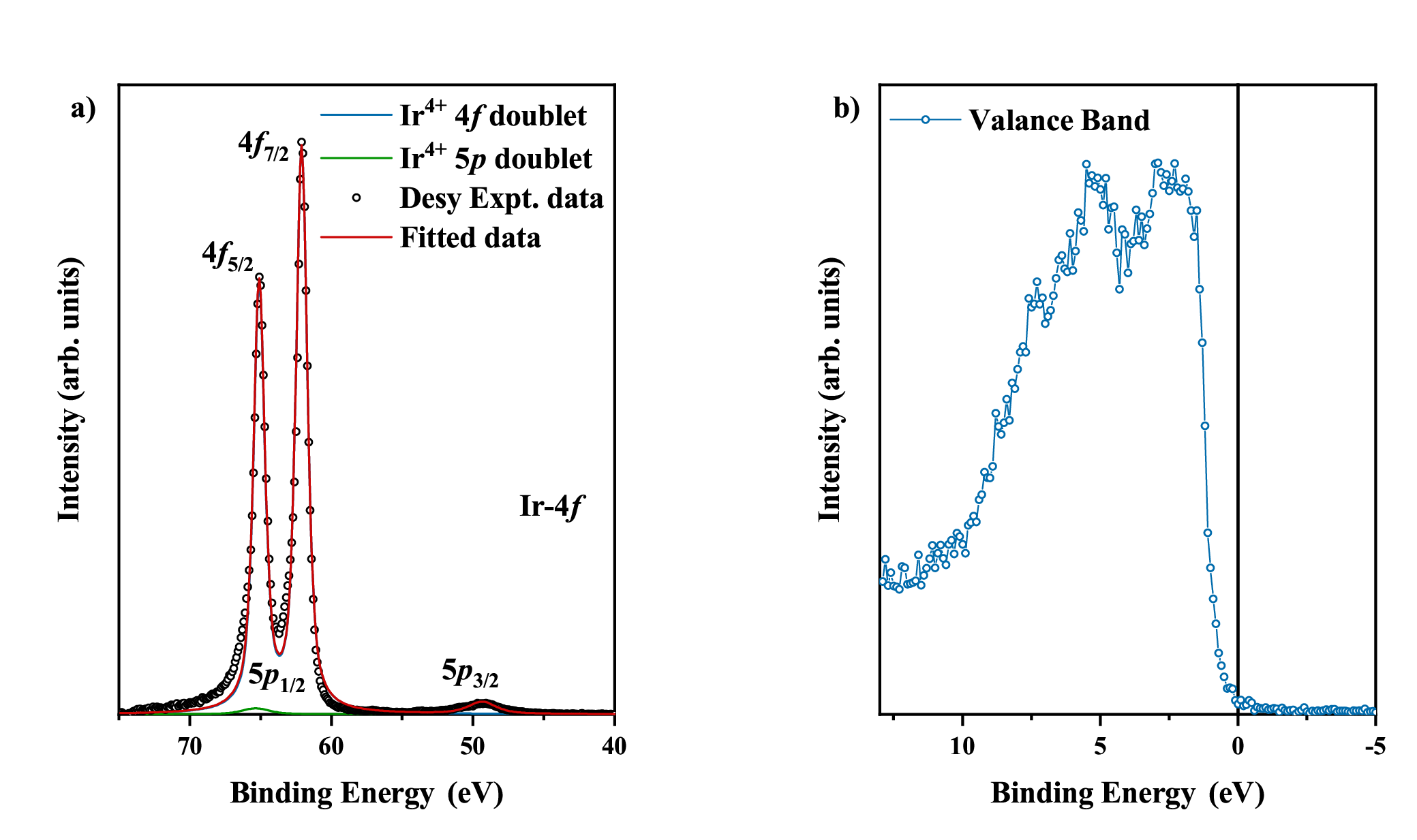}
\caption{a) Ir 4$f$ core level XPS spectrum (shaded black open circle) along with the fitting (red solid line). b) XPS valence band spectra of the Na$_2$IrCl$_6$ system }
\label{fgr:XPS}
\end{figure}

\subsection{Oxidation state from XPS}

X-ray photoelectron spectroscopy has been used to extract information about the oxidation state of Ir in Na$_2$IrCl$_6$. The Ir 4$f$ core level spectrum has been collected and shown in Figure~\ref{fgr:XPS}(a). The spectrum could be fitted with one single doublet having spin-orbit separation of 2.95 eV which confirms the presence of Ir$^{4+}$ in the system. The small envelope around 67~eV could be a result of partial screening of the photoemission core hole by the Ir 5$d$ electrons~\cite{r65}.However, the rise of density of states at the Fermi level at room temperature in the XPS valence band spectrum, as seen in Figure~\ref{fgr:XPS}(b), confirms that Na$_2$IrCl$_6$ also falls into that category of robust insulators~\cite{r34}.

\subsection{Magnetic Properties}

\begin{figure}[!ht]
\centering
\includegraphics[width=0.56\columnwidth]{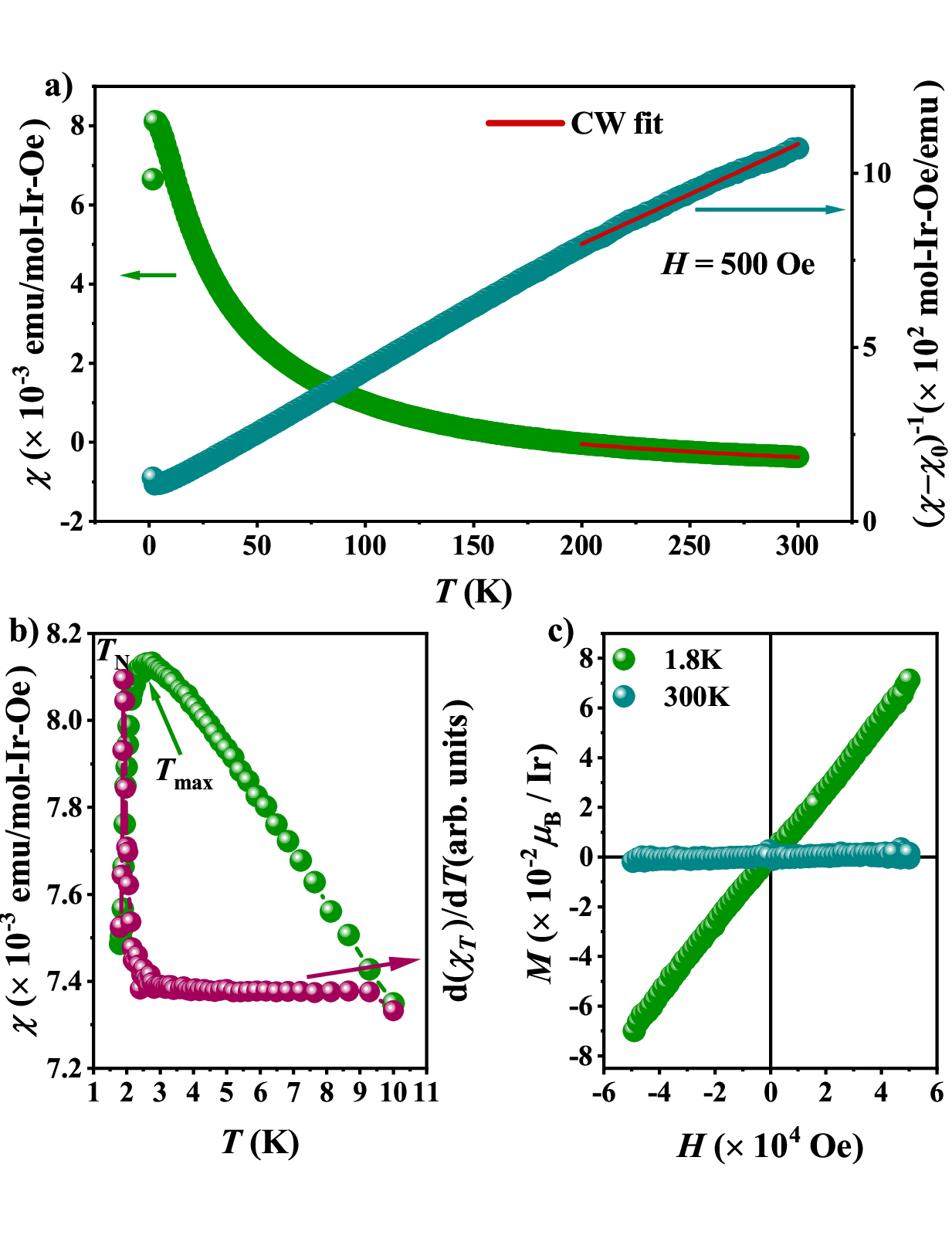}
\caption{Magnetic properties of Na$_2$IrCl$_6$. a) $\chi$ vs $T$ and ($\chi$-$\chi_0$)$^{-1}$ vs $T$ plots from 1.8 K to 300 K. b) $\chi$ vs $T$ and $\frac{d\chi(T)}{dT}$ vs $T$ plots from 1.8 K to 11 K. c) $M$ vs $H$ plots under 1.8 K and 300 K respectively.}
\label{fgr:Magnetic}
\end{figure}

\begin{figure}[!ht]
\centering
\includegraphics[width=0.87\textwidth]{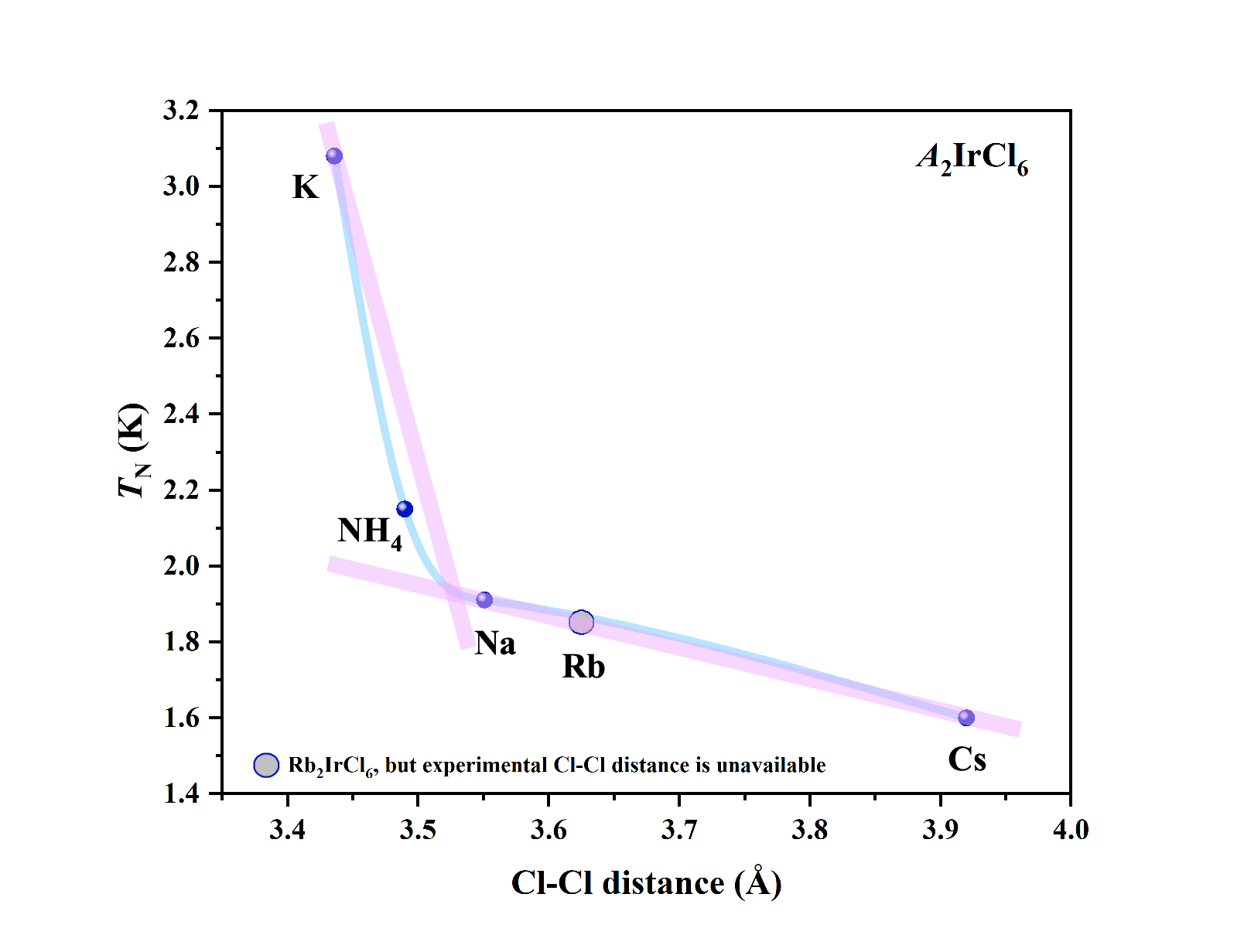}
\caption{ Transition temperatures of alkali iridium hexachlorides as function of nearest neighbour Cl-Cl distance}
\label{fgr:Cl-Cl distance}
\end{figure}

The temperature-dependent $dc$ magnetic susceptibility $\chi = \frac{M}{H}$ measured under the magnetic field applied $H$ = 500 Oe is shown in Figure~\ref{fgr:Magnetic}(a). Upon cooling, the susceptibility curve exhibits a broad maximum at $T_\text{max}$ $\approx$ 2.63 K implying the onset of short-range spin-spin correlations. The susceptibility data above 200 K have been fitted using the following expression,
\begin{equation}
\chi = \chi_0 + \frac{C}{T - \theta_\text{CW}}   
\end{equation}
where $\chi_0$ is the temperature-independent contribution due to the core diamagnetism ($\chi_\text{core}$) and Van Vleck paramagnetism ($\chi_\text{vv}$)~\cite{r34}. The second term represents the Curie-Weiss law with the Curie constant $C$ and Curie-Weiss temperature $\theta_\text{CW}$. The Curie constant is given by $C = \frac{N_\text{A}\mu_\text{eff}^2}{3k_\text{B}}$, where $N_\text{A}$ is Avogadro’s number, $\mu_\text{eff}$ is the effective magnetic moment, and $k_\text{B}$ is the Boltzmann constant. The least-square fitting with Eq. (1) above 200 K reveals $\chi_0$ = -1.3 × 10$^{-3}$ emu/mol, $C$ = 0.365 emu K/mol, and $\theta_\text{CW}$ = -85.2 K. The negative value of $\theta_\text{CW}$ implies the predominant AFM exchange interaction between the Ir$^{4+}$ ions in Na$_2$IrCl$_6$. Figure~\ref{fgr:Magnetic}(b) shows an expanded version of the susceptibility data between 1.8 to 10~K and the corresponding $\frac{d\chi(T)}{dT}$ plot capturing the sharp drop in susceptibility below $T_\text{max}$. As this $\frac{d\chi(T)}{dT}$ plot in Figure~\ref{fgr:Magnetic}(b) shows, the system orders antiferromagnetically at 1.91~K. The $M$($H$) data shown in Figure~\ref{fgr:Magnetic}(c) confirms the robust antiferromagnetism in the system as no hysteresis is observed down to 1.8 K. The frustration parameter, $f = |\frac{\theta_\text{CW}}{T_\text{N}}|$, is estimated to be quite large (44.6) suggesting presence of strong magnetic frustration in Na$_2$IrCl$_6$. The effective magnetic moment, $\mu_\text{eff}$ = 1.71 $\mu_\text{B}$, is in excellent agreement with the expected value $\mu_\text{eff} = g\sqrt{J(J + 1)}\mu_\text{B}$ using the Land\'e g-factor $g$ = 2 for the ideal $J_\text{eff}$ = $\frac{1}{2}$ state expected for Ir$^{4+}$ in the cubic crystal field~\cite{r34}. The temperature-independent contribution $\chi_0$~=~-1.83 × 10$^{-4}$ emu/mol includes the Van Vleck part $\chi_\text{vv}$ and the core part $\chi_\text{core}$, and the $\chi_\text{vv}$ is then evaluated by subtracting $\chi_\text{core}$ from $\chi_0$, resulting in a value of -1.1 × 10$^{-3}$ emu/mol. \\
All the above results show that for the first time, the long-time missing fcc antiferromagnetic VODP Na$_2$IrCl$_6$ single crystals could be produced which fits perfectly within the $A_2$IrCl$_6$ family, known for more than at least six decades, both in terms of their crystallographic and magnetic characteristics. It has been mentioned earlier that the AFM exchange couplings in hexachloroiridates are mediated through the Ir-Cl-Cl-Ir superexchange pathways and depends critically on the Cl-Cl superexchange distance~\cite{r32,r48,r49,r50,r51}. In Figure~\ref{fgr:Cl-Cl distance}, the antiferromagnetic transition temperatures of the various members have been plotted as a function of the corresponding Cl-Cl distance. It should be noted here that the Cl-Cl experimental distance is not known for the Rb$_2$IrCl$_6$ while only a calculated value along with transition temperature has been reported~\cite{r49}, and in Fig.~5 only these numbers have been summarized. It is clear from the figure that albeit nonmonotonically, the AFM exchange interaction strength and consequently the value of $T_N$ categorically reduces with increasing Cl-Cl distance, and at least for the initial members K, Na, and (NH$_4$), the dependence is almost linear.

\subsection{Electronic Structure \& Microscopic Magnetic Model}

\begin{figure}[ht]
	\begin{center}
		\includegraphics[width=0.5\columnwidth]{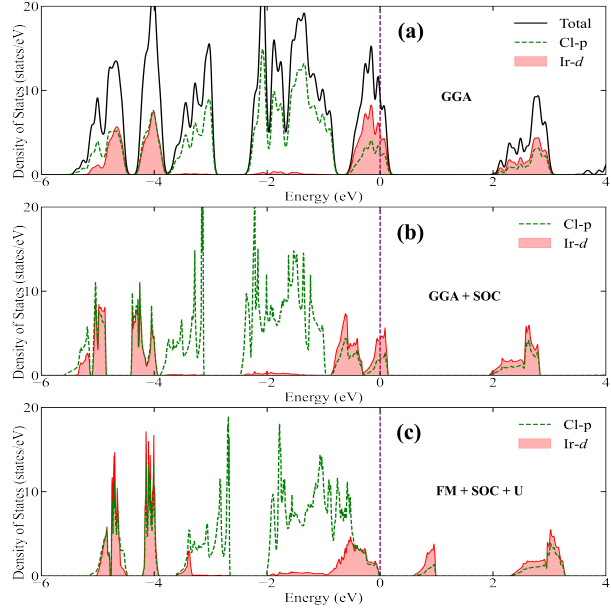}
		\caption{Atomic and orbital-resolved DOS from the non-magnetic GGA (top) GGA+SOC (middle) and ferromagnetic GGA+SOC+U (bottom) calculations with $U_d=1.7$ eV and $J_H=0.3$ eV.}
		\label{dos}
	\end{center}
\end{figure}

We now proceed to the computational analysis. The non-spin polarized GGA density of states shown in Fig.\ref{dos}(a) is consistent with the nominal ionic formula, Na$_2$$^{2+}$Ir$^{4+}$Cl$_6$$^{6-}$. As a result the fermi level is dominated by partially filled Ir-$d$ states and completely filled Cl-$p$ states. The Ir atoms is in the octahedral environment, generated by the surrounding Cl ligands, which splits the $d$ states into lower lying $t_{2g}$ states and higher $e_g$ states. The crystal field splitting is clearly visible in the nonmagnetic GGA density of states, where the partially filled $t_{2g}$ states are at the Fermi level, while the unoccupied $e_g$ states lie at considerably higher energy (see Fig.\ref{dos}(a)). Nevertheless, the projected density of states shown in Fig.\ref{dos} reveals substantial Cl $3p$ weight in the same energy range as the Ir $5d$ states, indicating appreciable Ir-Cl hybridization.

Within the $t_{2g}$ sector, SOC reorganizes the bands into predominantly
$j_{\mathrm{eff}}$~=~$\frac{3}{2}$ and $j_{\mathrm{eff}}$~=~$\frac{1}{2}$ manifolds. In the ideal cubic limit, the
$j_{\mathrm{eff}}$~=~$\frac{3}{2}$ quartet is nominally filled, whereas the $j_{\mathrm{eff}}$~=~$\frac{1}{2}$ doublet is
half filled for Ir$^{4+}$. The GGA+SOC density of states displayed in Fig.\ref{dos}(b) shows that the Ir- $t_{2g}$ manifold, extending from approximately -1.0 to 0.3 eV, develops a characteristic two-peak structure predominantly associated with the $j_{\mathrm{eff}}$~=~$\frac{3}{2}$ and $j_{\mathrm{eff}}$~=~$\frac{1}{2}$ states (middle panel of Fig.\ref{dos}). Incorporating Coulomb correlation in ferromagnetic configuration in our calculation
opens up a band gap of aprroximately 0.6 eV. The spin and orbital moments obtained from our calculations are consistent with the anticipated $J_{\mathrm{eff}}$~=~$\frac{1}{2}$ state of Ir$^{4+}$. We find the spin moment to be 0.33 ${\mu}_\text{B}$ and orbital moment to be 0.6 ${\mu}_\text{B}$ per Ir atom. These results establish the correlated, SOC-assisted insulating character of \NaIrCl{} and define the low-energy subspace used below to construct the magnetic Hamiltonian.

The low-energy magnetic degrees of freedom are represented by pseudospin $S$~=~$\frac{1}{2}$ moments centered on the Ir sites. Exchange couplings in Ir$^{4+}$ compounds are generally anisotropic. The most general bilinear spin Hamiltonian can be written as
\begin{equation}
  \mathcal{H}
  =\sum_{\langle ij\rangle}
   \mathbf{S}_i^{\mathsf T}\boldsymbol{\mathcal{J}}_{ij}\mathbf{S}_j
  = \sum_{\langle ij\rangle}
    \sum_{\alpha,\beta=x,y,z}
    S_i^{\alpha}\mathcal{J}_{ij}^{\alpha\beta}S_j^{\beta},
  \label{eq:general_hamiltonian}
\end{equation}
where $i, j$ label the Ir sites, whereas $\alpha, \beta$ label the Cartesian spin component. The quantity $\boldsymbol{\mathcal{J}}_{ij}$ is the complete $3\times3$ exchange
matrix associated with the bond connecting sites $i$ and $j$, while
$\mathcal{J}_{ij}^{\alpha\beta}$ denotes one element of this matrix. In general, the exchange matrix can be decomposed into isotropic, antisymmetric, and symmetric anisotropic contributions:
\begin{equation} 
	\boldsymbol{\mathcal{J}}_{ij}
	=
	\begin{pmatrix}
		J_{ij}^{\mathrm{iso}}+\Lambda_{ij}^{xx}
		&
		D_{ij}^{z}+\Lambda_{ij}^{xy}
		&
		-D_{ij}^{y}+\Lambda_{ij}^{xz}
		\\[2mm]
		-D_{ij}^{z}+\Lambda_{ij}^{xy}
		&
		J_{ij}^{\mathrm{iso}}+\Lambda_{ij}^{yy}
		&
		D_{ij}^{x}+\Lambda_{ij}^{yz}
		\\[2mm]
		D_{ij}^{y}+\Lambda_{ij}^{xz}
		&
		-D_{ij}^{x}+\Lambda_{ij}^{yz}
		&
		J_{ij}^{\mathrm{iso}}+\Lambda_{ij}^{zz}
	\end{pmatrix}. \label{eq:exchange_tensor_general} 	
\end{equation}
Here, $J_{ij}^{\mathrm{iso}}$ is the isotropic Heisenberg exchange on the bond $\langle ij\rangle$, $\mathbf{D}_{ij}$ is the Dzyaloshinskii--Moriya vector, and $\boldsymbol{\Lambda}_{ij}$ is the symmetric part of the anisotropic exchange, satisfying $\Lambda_{ij}^{\alpha\beta}=\Lambda_{ij}^{\beta\alpha}$. For the bonds considered here, the Dzyaloshinskii--Moriya interaction vanishes because of inversion symmetry at the bond center, so that $\mathbf{D}_{ij}=0$.

\begin{figure}[ht]
	\begin{center}
		\includegraphics[width=0.5\columnwidth]{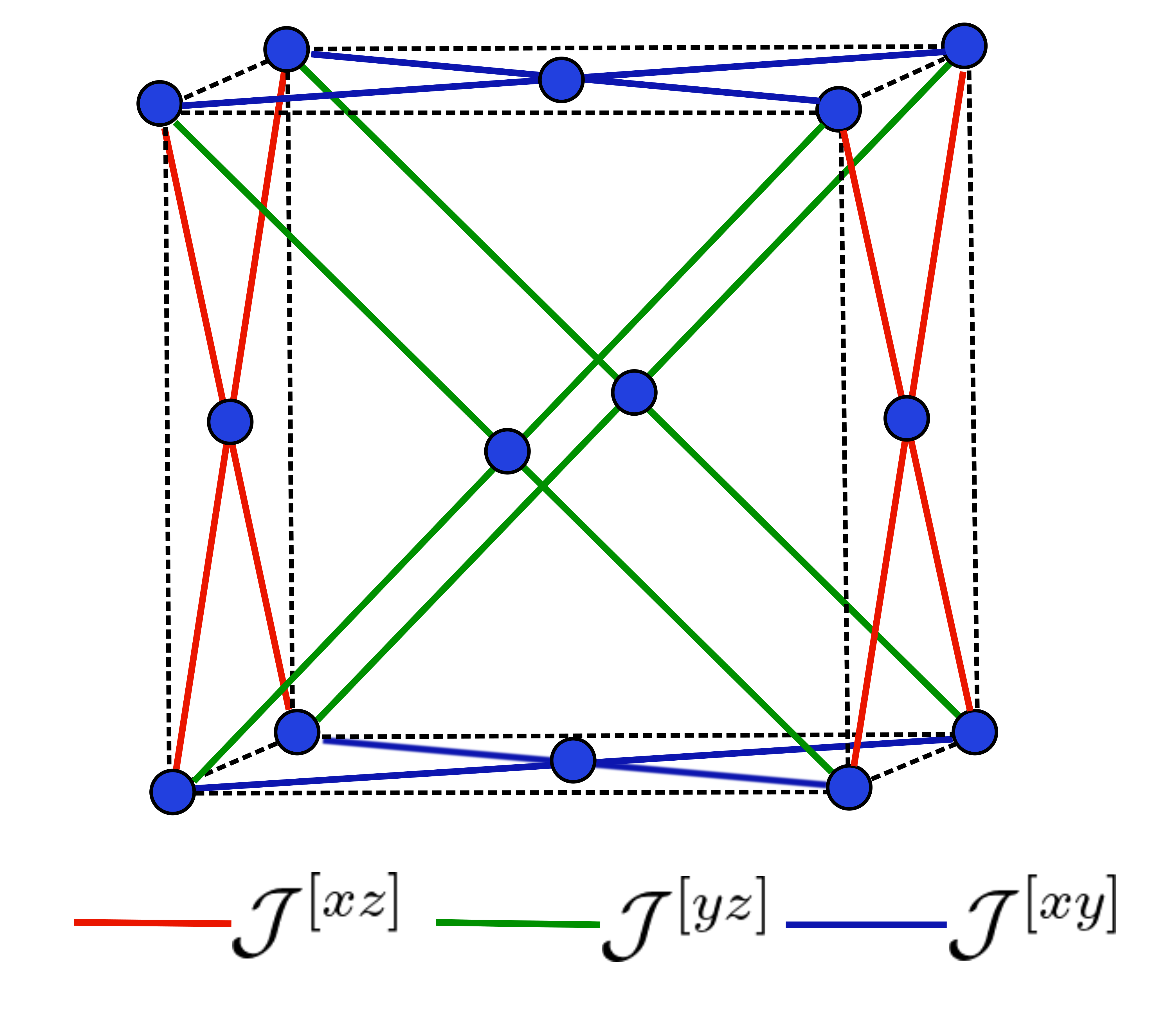}
		\caption{fcc spin lattice with different anisotropic interactions of the $J-K-\Gamma$ model on different faces of the cube.}
		\label{exch_path}
	\end{center}
\end{figure}

For the cubic structure, the twelve nearest-neighbour bonds of the fcc Ir sublattice can be divided into three families according to the plane containing the bond vector (see Fig.\ref{exch_path})\cite{Tsirlin_2019}. We denote the family of the bond connecting sites $i$ and $j$ by
\begin{equation}
	b_{ij}\in\{xy,xz,yz\}.
\end{equation}
The exchange matrix on a particular bond can therefore be written as
\begin{equation}
	\boldsymbol{\mathcal{J}}_{ij}
	=
	\boldsymbol{\mathcal{J}}_{\eta_{ij}}^{[b_{ij}]},
	\label{eq:bond_dependent_exchange}
\end{equation}
where $\eta_{ij}=\pm1$ determines the bond-orientation-dependent sign of the off-diagonal exchange interaction. In the global cubic coordinate system, the three bond-family matrices are

\begin{equation}
	\boldsymbol{\mathcal{J}}_{\eta}^{[xy]}
	=
	\begin{pmatrix}
		J & \eta\Gamma & 0\\
		\eta\Gamma & J & 0\\
		0 & 0 & J+K
	\end{pmatrix},
	\qquad
	\boldsymbol{\mathcal{J}}_{\eta}^{[xz]}
	=
	\begin{pmatrix}
		J & 0 & \eta\Gamma\\
		0 & J+K & 0\\
		\eta\Gamma & 0 & J
	\end{pmatrix},
	\qquad
	\boldsymbol{\mathcal{J}}_{\eta}^{[yz]}
	=
	\begin{pmatrix}
		J+K & 0 & 0\\
		0 & J & \eta\Gamma\\
		0 & \eta\Gamma & J
	\end{pmatrix}.
	\label{eq:bond_family_matrices}
\end{equation}
Here, $J$ is the isotropic Heisenberg exchange coupling, $K$ is the bond-dependent Kitaev exchange coupling, and $\Gamma$ is the symmetric off-diagonal exchange coupling. The labels $[xy]$, $[xz]$, and $[yz]$ denote the complete bond families and should not be confused with the Cartesian components $\mathcal{J}_{ij}^{xy}$, $\mathcal{J}_{ij}^{xz}$, and
$\mathcal{J}_{ij}^{yz}$ of the exchange matrix.

The exchange parameters $J$, $K$ and $\Gamma$ were determined using
exact diagonalization of a two-site model \cite{Valenti_2016, Dasgupta_2026, Das2025}. The nearest-neighbour $t_{2g}$ hopping matrix for a bond belonging to the $xy$ family can
be written in the $(d_{yz},d_{xz},d_{xy})$ basis as
\begin{equation}
	\boldsymbol{T}^{[xy]}
	=
	\begin{pmatrix}
		t_1 & t_2 & t_4\\
		t_2 & t_1 & t_4\\
		t_4 & t_4 & t_3
	\end{pmatrix}.
	\label{eq:xy_hopping_matrix}
\end{equation}

From the DFT-derived Wannier Hamiltonian, we obtained the nearest-neighbor hopping matrix, with hopping parameters $t_1=2.2$ meV, $t_2=-9.3$ meV, $t_3=-77.1$ meV, and $t_4=0.0$ meV. The corresponding two-site exact-diagonalization calculation using $\lambda$ = 0.4 eV, $J_H$ = 0.3 eV, and $U_d$ = 1.7 eV yields the exchange parameters $J=1.7$ meV, $K=0.6$ meV and $\Gamma=-0.1$ meV. 

Independently, using identical values for $\lambda$, $J_H$ and $U_d$, we evaluated the nearest-neighbor exchange parameters within second-order perturbation theory derived for the limit $U \gg \lambda \gg t$ \cite{Valenti_2016}. The resulting values coincide with those obtained from the two-site exact-diagonalization calculation, providing a consistent description of the microscopic magnetic interactions for a $J_{\mathrm{eff}}$~=~$\frac{1}{2}$ system. Throughout this work, positive exchange constants denote
antiferromagnetic interactions.  The ratios $K/J=$ 0.35,
$\Gamma/J=$ -0.06 show that \NaIrCl{} is placed into the collinear antiferromagnetic region in the classical phase diagram for the $J-K-\Gamma$ model on fcc lattice \cite{Paramekanti_2015}. The calculated Curie--Weiss temperature in the adopted convention is
\begin{equation}
  \Theta_{\mathrm{CW}}^{\mathrm{calc}}
  = -\frac{S(S+1)}{3k_{\mathrm B}}
    \sum_j \frac{1}{3}\operatorname{Tr}(\mathsf{J}_{ij})
  = -66.1 K,
  \label{eq:theta_cw}
\end{equation}
which may be compared directly with the experimental value
$\Theta_{\mathrm{CW}}^{\mathrm{exp}}=$ -85 K. The second-neighbor exchange obtained from perturbation theory is $J_2=0.012~\mathrm{meV}$, which is below the reliable numerical resolution of the calculation and may therefore be regarded as negligible. Such a small value is expected owing to the large second-neighbor Ir--Ir separation of (9.863~\text{\AA}), which strongly suppresses the corresponding hopping and exchange interaction.

\section{CONCLUSIONS}

In summary, for the first time we have successfully grown the purely anhydrous, {\it face-centered cubic} antiferromagnetic vacancy ordered double perovskite Na$_2$IrCl$_6$, which remained missing for more than six decades now. Slow evaporation technique at low temperatures helped us to achieve this. Detailed crystallographic, magnetic, and other characterizations of the crystals revealed that Na$_2$IrCl$_6$ is a new fcc antiferromagnet with somewhat enhanced Cl-Cl distance as well as lattice parameter and consequently possess  somewhat lower antiferromagnetic transition temperature. Interestingly the [IrCl$_6$]$^{2-}$ octahedra appears completely undistorted and therefore, may come out to be the next virgin exploration ground for studying the `perfect' $J_{\mathrm{eff}}$~=~$\frac{1}{2}$ ground state and also to probe the applicability of the ideal Heisenberg-Kitaev model in this class of systems.

\section*{ACKNOWLEDGMENTS}
S.S.S. , A.P. and T.C. acknowledge Indian Association for the Cultivation of Science (IACS) and Department of Science and Technology (DST), INDIA for the fellowship. The authors also thank the Central Scientific Services (CSS) and the Technical Research Center (TRC) of IACS for providing experimental facilities. The authors acknowledge the financial support for XPS experiments at the DESY synchrotron radiation facility (I-20251160). S.R. thanks the Science and Engineering Research Board (SERB) under the Department of Science and Technology (DST), India, for financial support (Project No. CRG/2023/001060).

\printbibliography

\end{document}